\documentclass[11pt]{article}
\usepackage{graphicx}

\newsavebox{\PSLASH}
\sbox{\PSLASH}{$p$\hspace{-1.8mm}/}

\begin{document}
\title{Boundary conformal field theories and loop models  }
\author{M. A. Rajabpour$^{a,b}$\footnote{e-mail: rajabpour@to.infn.it} \\ \\
  $^{a}$Dip. di Fisica Teorica and INFN, Universit\`{a} di
Torino, Via P. Giuria 1, 10125 Torino,
\\Italy\\$^{b}$Institute for Studies in Theoretical Physics and Mathematics,
Tehran 19395-5531, Iran} \maketitle
\begin{abstract}
We propose a systematic method to extract conformal loop models for
rational conformal field theories (CFT). Method is based on defining
an ADE model for boundary primary operators by using the fusion
matrices of these operators as  adjacency matrices. These loop
models respect the conformal boundary conditions. We discuss the
loop models that can be extracted by this method for minimal CFTs
and then we will give dilute $O(n)$ loop models on the square
lattice as examples for these loop models. We give also some
proposals for WZW $SU(2)$ models.
  \vspace{5mm}%
\end{abstract}
\section{Introduction}\
The study of statistical models related to loop models is
interesting both from the physical and the mathematical point of
views. Most of the statistical  models studied in physics such as
the Ising, the  q-state Potts model and also complicated vertex
models can be represented in terms of loops \cite{Nienhis1}. The
loop representation of the spin system is very easy to understand:
loops correspond to domain walls separating regions of different
magnetization. The study of critical loop models can be interesting
from many point of views: they are good candidates for the ground
state of topological quantum systems \cite{freedman}, they are also
good candidates for the Schramm Loewner evolution (SLE), a method
discovered by Schramm \cite{schramm} to classify  conformally
invariant curves connecting two distinct boundary points in  a
simply connected domain.

Different applications of conformal loop models are stimulating to
do a systematic study of these models by CFT. Recently we proposed
in ~\cite{Rajabpour} a method to extract loop models corresponding
to a conformal field theory (CFT), the method was based on defining
a RSOS model for every primary operator by using fusion matrix of
the primary operator as an adjacency matrix and then extracting the
loop model corresponding to domain walls of the RSOS model. The
weight of the loop model is equal to the quantum dimension of the
corresponding operator. In this paper we want to follow the same
method consistent with the conformal boundary operators, since the
SLE is a boundary CFT we think that using the fusion matrix of
boundary operators as an adjacency matrix is more consistent with
the nature of SLE. Recently a very nice and strong project was
initiated by Jacobsen and Saleur \cite{Jacobsen and Saleur1}
followed by Dubail, Jacobsen, Saleur \cite{Jacobsen and Saleur2} to
classify all the possible conformal boundary loop models. It is
based on classifying the possible boundary loop models compatible
with the boundary conformal field theories. This classification is
in close relation with the earlier work by Cardy on formulating the
modular invariant partition function of $O(n)$ model on the annulus
\cite{Cardy1}. The results that we get by our method apart from
simplicity are all compatible with the results in \cite{Jacobsen and
Saleur1,Jacobsen and Saleur2,Cardy1}.

The paper is organized as follows: In the next section we will
introduce the necessary ingredients to find the boundary operators
and also the fusion matrices corresponding to them. In the third
section we briefly review the method proposed in ~\cite{Rajabpour}
and we will also generalize it to the graphs with largest eigenvalue
bigger than two. The central claim of this section is as follows:
the loop model extracted with this method is connected with the
properties of the statistical loop model in the same universality
class as the corresponding CFT. In the third section we follow
explicitly some examples in particular; Ising model, tri-critical
Ising model, three states Potts model and tri-critical three states
Potts model. Then we will give the possible loop models, extractable
with this method, of minimal CFTs and also the lattice models
corresponding to these loop models. We will close this section by
giving some proposals for possible loop models for WZW $SU(2)$
models. Last section contains our conclusions with a brief
description of the work in progress motivated by these results.

\section{Boundary conformal field Theory}\
\setcounter{equation}{0}

To define loop model for a generic minimal CFT consistent with the
conformal boundary we need to first summarize the main important
facts about boundary CFT. The most important ingredient to classify
the boundary conformal operators is the modular invariant partition
function of the CFT. The classification of modular invariant
partition functions of $SU(2)$ minimal models are well known and can
be related to a pair of simply laced Dynkin diagrams $(A,G)$
~\cite{CapItzZub87}. The complete classification based on ADE
diagrams is
\begin{eqnarray}\label{ADE} (A,G)=\cases{ (A_{h-1},A_{g-1})&\cr
(A_{h-1},D_{(g+2)/2}),\quad g\ \mbox{even}&\cr (A_{h-1},E_6),
\quad\quad\quad\hspace{0.2cm} g=12 &\cr (A_{h-1},E_7),
\quad\quad\quad\hspace{0.2cm} g=18&\cr (A_{h-1},E_8),
\hspace{0.2cm}\quad\quad\quad g=30,}
\end{eqnarray}
where $g$ and $h$ are the Coxeter numbers of $A$ and $G$ with
$h,g\geq 2$. The above pair of Dynkin diagrams  describes bulk
modular invariant partition function with some primary operators and
with the following central charge
\begin{eqnarray}\label{central charge}
 c=1-6\frac{(h-g)^2}{h g}.
\end{eqnarray}
Each of the unitary minimal models $M(A_{h-1},G)$ with $g-h=\pm1$
can be realized as the continuum scaling limit of an integrable
two-dimensional lattice model at criticality, with heights living on
the nodes of the graph $G$. In particular, the critical series with
$g-h=1$ is associated with the A-D-E lattice models~\cite{ABF84} and
the tri-critical series with $g-h=-1$ is associated with the dilute
lattice models ~\cite{Roc92,WN}. For theories with a diagonal torus
partition function it is known that there is a conformal boundary
condition associated to each operator in the theory ~\cite{Cardy89}.
The fusion rules of these boundary operators are just given by the
bulk fusion algebra. It was shown in a series of papers that for
$SU(2)$ minimal models one can propose a complete set of conformal
boundary operators $i=(r,a)\in (A,G)$, where $r$ and $a$ are nodes
on the Dynkin diagram of $A$ and $G$ respectively with the
identification $(r,a)=(h-r,\gamma(a))$, where $\gamma$ is an
automorphism acting on the nodes of the graph $G$. This automorphism
is identity except for the $A$, $E_{6}$ and $D_{odd}$ which is
$Z_{2}$ symmetry of Dynkin diagram, symmetries of Dynkin diagrams
play an important rule in the forthcoming discussion. Following
\cite{BPZ} we show the corresponding operators by $\hat{\phi}_{i}$
and the independent boundary states by $|(r,a)\rangle$ which is
called Cardy states. Cardy states can be written in terms of
Ishibashi states, i.e. $|j\rangle\rangle$, as follows
$|(r,a)\rangle=\sum_{j} c_{(r,a)}^{j}|j\rangle\rangle$, where sum is
over all Ishibashi states. We are interested to the fusion rules of
these boundary operators. To give a formula for the fusion rules of
these operators we need to define some quantities. Let $\Psi$ be the
eigenvectors of the adjacency matrix corresponding to the group $G$
then the graph fusion matrices $\hat{N}_{a}$ with $a\in G$ can be
defined as follows
\begin{eqnarray}\label{graph fusion matrix}
(\hat{N}_a)_b{}^c= \sum_{m\in \mbox{\scriptsize Exp}(G)} {\Psi_{am}
\Psi_{bm} \Psi^*_{cm}\over \Psi_{1m}},\qquad a,b,c\in G,
\end{eqnarray}
where $Exp(G)$ denotes the set of exponents of $G$, see table ~1.
Let's show also the graph fusion matrix for $A_{h-1}$ by $N_{r}$
then following \cite{BPZ} the fusion rules for boundary operators
are
\begin{eqnarray}\label{fusion rules}
\hat{\phi}_{i_{1}}\hat{\phi}_{i_{2}}=\sum_{i_3\in
(A,G)}(\mathcal{N}_{i_{1}})_{i_{2}}^{i_3}\hat{\phi}_{i_{3}},
\end{eqnarray}
where $(\mathcal{N}_{i_{1}})_{i_{2}}^{i_3}$ has the following
relation with the graph fusion matrices of $A$ and $G$
\begin{eqnarray}\label{fusion rules2}
(\mathcal{N}_{(r_{1},a_{1})})_{(r_{2},a_{2})}^{(r_{3},a_{3})}=N_{r_{1}r_{2}}^{r_{3}}\hat{N}_{a_{1}a_{2}}^{a_{3}}.
\end{eqnarray}
For more details about the connection of the boundary operators to
bulk counterparts see ~\cite{BPZ, BpPZ}.
\newline
\begin{table}[htp]

\begin{center}
\begin{tabular}{|c|c|c|}\hline
Dynkin Diagram &  Coexter Number($h$)  & Coexter Exponent($m$) \\
\hline

$A_{n}$ &$n+1$ &$1,2,...,n$    \\
\hline

$D_{n}$ &$2(n+1)$ & $1,3,...,2n-1,n-1$
  \\ \hline

$E_{6}$ &$12$ & $1,4,5,7,8,11$
\\ \hline
$E_{7}$&$18$ & $1,5,7,9,11,13,17$
\\ \hline
$E_{8}$ &$30$ & $1,7,11,13,17,19,23,29$\\
\hline
\end{tabular}
\end{center}
\caption{\label{tab1} The Coexter number and the Coexter exponents
of Dynkin diagrams. }
\end{table}
To calculate the fusion matrices of boundary operators we need also
to define a conjugation operator $C(a)=a^*$, it is the identity
except for $D_{4n}$ graphs where the eigenvectors
 $\Psi_{am}$ are complex and conjugation
corresponds to the $Z_2$ Dynkin diagram automorphism. It then
follows that $\hat{N}_{a^*b}^c=\hat{N}_{ca}^b$. The operator $C(a)$
acts on the right to raise and lower indices in the fusion matrices
$\hat{N}^a=\hat{N}_a C$ so it is the important ingredient to get the
right fusion matrices for the boundary operators, in particular for
the $D_{4n}$ graphs. we will give some examples in section ~4, in
particular we use the above method to get the fusion matrices of the
boundary operators of Ising model, tri-critical Ising model, ~3
state Potts model and tri-critical ~3 state Potts model.

\section{Loop Models for Boundary operators}\
\setcounter{equation}{0}

In this section we propose a method to extract some possible loop
models for CFTs, the method is the same as the method introduced
recently in ~\cite{Rajabpour}. In that reference we showed that
using the fusion matrix as an adjacency matrix it is possible to
associate a $O(n)$ loop model to every primary operator. The method
is briefly as follows: The graph of a primary operator
$\hat{\phi}_{i}$ has $g$ vertices where $g$ is the number of primary
operators in the theory and edges connecting pairs of vertices
$(j,k)$ when $\mathcal{N}_{ij}^{k}=1$. Following \cite{Cardy ADE}
one can define a height model on the triangular lattice by imposing
that the height $h_{j}$ at the site $j$ can take values
$0,1,\dots,g-1$. Then constraint the heights at neighboring sites
according to the incidence matrix associated to a given primary
field $\hat{\phi}_{i}$: only neighbor heights $h_{j}$ and $h_{k}$
with $(\mathcal{N}_{i})_{j}^{k}=1$ are admissible. For a consistent
definition of loop models on a triangular lattice at least two of
the heights at the corners of  an elementary triangular plaquette
should be equal then the weights for the elementary plaquette are
defined  as follows. If the  heights of plaquette are $(c,b,b)$ with
$c \ne b$ then the weight is
$x(\frac{\hat{S}_{l}^{b}}{\hat{S}_{l}^{c}})^{1/6}$, where $\hat{S}$
satisfies
$\sum_{b}(\mathcal{N}_{a})_{b}^{c}\frac{\hat{S}_{l}^{b}}{\hat{S}_{l}^{0}}=\frac{\hat{S}_{l}^{a}}{\hat{S}_{0}^{l}}\frac{\hat{S}_{l}^{c}}{\hat{S}_{l}^{0}}$.
It means that the $b$ th element of the eigenvector of
$\mathcal{N}_{a}$ with eigenvalue
$\frac{\hat{S}_{l}^{a}}{\hat{S}_{0}^{l}}$ is given by
$\frac{\hat{S}_{l}^{b}}{\hat{S}_{l}^{0}}$.  If the heights are all
equal then the weight is $1$ except for those with
$\mathcal{N}_{ab}^{b}\neq 0$ that have weights  $1$ or $x$ depending
on the particular model considered \footnote{For more details
specially about identical neighbor heights see \cite{Rajabpour}.}.
The next step is to mark triangles with unequal heights $(c,b,b)$
drawing a curved segment on the dual honeycomb lattice \cite{Cardy
ADE} and linking to the center the midpoints of the two  edges with
different heights ($b$ and $c$) at the extremes (See fig ~1).
Summing over the admissible values of heights consistent with a
given loop configuration we find
\begin{eqnarray}\label{loop height}
\sum_{b}(\mathcal{N}_{a})_{b}^{c}\frac{\hat{S}_{l}^{b}}{\hat{S}_{l}^{c}}=\frac{\hat{S}_{l}^{a}}{\hat{S}_{0}^{l}},
\end{eqnarray}
where the sum is just over $b$. We  take most of the times $l=0$ to
get the largest eigenvalue of $\mathcal{N}_{a}$ to guaranty positive
real weights in our height models, however, we will also point to
other cases.
\begin{figure}
\begin{center}
\includegraphics[angle=0,scale=0.6]{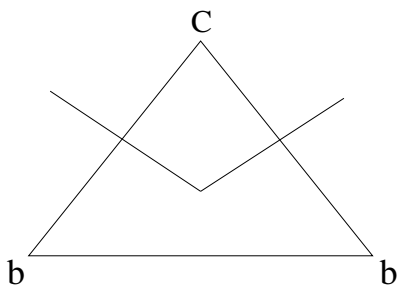}
\caption{A triangular plaquette  with $c\neq b$ and the
corresponding curve segment on the dual honeycomb lattice.}
\label{Fig1}
 \end{center}
\end{figure}
The weight of the loops is given by the largest eigenvalue of the
fusion matrix $n_{a}$ and the partition function of the model is as
follows
\begin{eqnarray}\label{O(n)}
Z=\sum x^{l}n_{a}^{N},
\end{eqnarray}
where $l$ is the number of bonds in the loop configuration and $N$
is the number of loops. Using this method we can correspond to every
boundary conformal operator a $O(n)$ loop model, since the $O(n)$
model posses a dilute critical point for $n\leq 2$ with
$x_{c}=\frac{1}{\sqrt{2+\sqrt{2-n}}}$; see ~\cite{Nienhis2},
correspondingly our loop models will have a critical point just for
the fields with $n_{a}$ smaller than $2$. The $O(n)$ model has
another critical regime, the so-called dense phase, for
$x=(x_{c},\infty)$ corresponds to a different universality class.
Mapping to the $O(n)$ model helps us to find the connection with
SLE: from coulomb gas arguments we know that, in the dilute regime,
the loop weight has the following relation with the drift in the SLE
equation
\begin{eqnarray}\label{S minimal}
n_{a}=-2 \cos(\frac{4\pi}{\kappa}).
\end{eqnarray}
For the dense phase the above equation is still true if we work in
the region $4\leq \kappa \leq 8$. Using the above equation we can
find the properties of the loop model corresponding to a boundary
conformal operator. The achievement of this method is respecting the
Cardy's equation \cite{Cardy89}: \textit{fields in the same sector
have the same loop representation}.

Before generalizing the definition to more general graphs we should
stress that although we started with well defined minimal CFT but
the loop model that we extracted is not necessarily minimal. The
point is that the extracted loop model respects some aspects of the
corresponding conformal field theory. This is like to say that
although the domain walls in Ising model at the critical point is
the same as the critical $O(n=1)$ but the Ising conformal field
theory does not explain all the aspects of the critical curves. From
this point the loop model that one can get by this method from the
rational CFT is not perfectly equal to the corresponding CFT.

One can generalize the above idea to the decomposable fusion graphs
by the method that was explained in \cite{Fendley}. Since the fusion
graphs of some operators in minimal models are equivalent to the
tensor product of two adjacency diagrams  one can use this method to
extract new loop models that can also have configurations with
crossing loop segments. The general strategy is based on extracting
critical loop models with $n\leq 2$ for the graphs with largest
eigenvalue bigger than ~2. Some graphs obey simple decomposition,
can be written as tensor product, but others need to be mapped to
simple decomposable graphs by going to the ground state adjacency
graph \cite{Fendley}. Here we just comment on decomposable graphs
$\mathcal{N}=\mathcal{N}_{1}\otimes \mathcal{N}_{1}$, where
$\mathcal{N}_{1}$ and $\mathcal{N}_{2}$ are simple ADE diagrams. In
these cases we can define two-flavor loop model living on the
honeycomb lattice independently, one is related to the loop model of
$\mathcal{N}_{1}$ with weight $n_{1}$ and the other comes from the
graph $\mathcal{N}_{2}$ with weight $n_{2}$. Fendley showed
\cite{Fendley} that in this case it is also possible to define
consistently interacting loop model on the square lattice with
partition function $Z=\sum n_{1}^{N_{1}}n_{2}^{N_{2}}b^{C}$, where
$N_{1,2}$ are the numbers of each kind of loop and $C$ is the number
of plaquettes with a resolved potential crossing at their center.
The critical values of $b$ were calculated in \cite{Fendley3} but
the critical properties of the loops are still unsolved. This is
obviously is not the only method to define loop model for non-simple
graphs, the other method is based on the multi-flavor loop model of
\cite{WN}. In this loop model a curve of flavor $i$ separating two
neighboring sites does not necessarily separate two sites with
different heights, for the definition of the RSOS model in this case
and its relation to the loop model see \cite{WN}.

In the next section we summarize some simple examples including the
most familiar minimal conformal models such as Ising, tri-critical
Ising, ~3-state Potts model and tri-critical ~3-state Potts model.
The main point is to take the fusion graphs as  adjacency graphs in
the consistent way and to extract some loop models. These loop
models are not equivalent to the corresponding conformal field
theory but still carry some aspects of the underlying field theory
in the consistent way, in particular the critical properties of
these loop models are in close connection with the corresponding
conformal field theory.

In this paper some distinctions are crucial. We have some minimal
conformal field theories with well defined fusion matrices and
modular invariant partition functions, one example is Ising
conformal field theory. There are some statistical models such as
spin models, RSOS models which at the critical point can be describe
partially by the minimal CFT, so the Ising CFT is different from the
statistical Ising model. We prefer also to distinguish between for
example dilute ADE models and dilute $O(n)$ loop model. They can be
mapped to each other and have the same phase transitions but since
the fundamental objects in one side is local and in the other one is
non-local this distinction is useful. There are lots of work done on
connecting these two models, minimal conformal field theories and
statistical models counterparts,  using integrability methods and
our argument hardly has something new to say from this point of
view. Finally we are defining another statistical model by using the
fusion matrices of primary operators of conformal field theory which
most of the times is in the same universality class as the
statistical model counterpart of the corresponding CFT. These height
models have also loop representations.  This similarity can be
useful to get an idea about the loop properties of the statistical
models with well-known minimal CFTs.

\section{Some Examples}\
In this section we apply the method introduced in section ~3 to the
minimal conformal field theories with well defined fusion structure
and also WZW $SU_{k}(2)$ models. We will also point on the
consistency of these loop models with the Cardy's boundary states.
These consistency is a hint to believe that it may be possible to
extend the results in to the level of the boundary partition
function \cite{Cardy1}. For notational convenience in this section
of the paper we will drop the hat of boundary operators.
\newline
\begin{figure}
\begin{center}
\includegraphics[angle=0,scale=0.8]{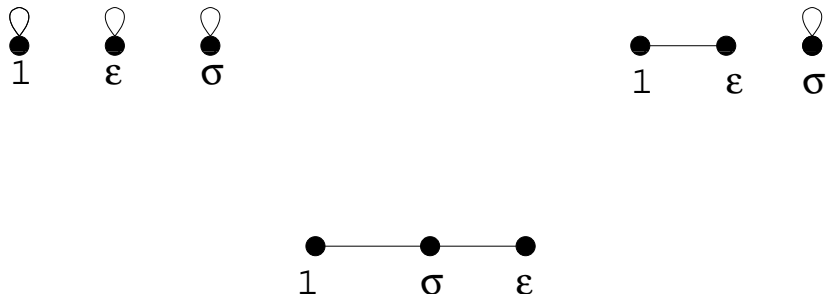}\\
\caption{ Graphs of fusion matrices of boundary primary operators in
Ising model, from the left to the right the fusion graphs of $1$,
$\epsilon$ and $\sigma$. The graph of the operator $\epsilon$ is
$A_{2}$ and the graph of $\sigma$ is $A_{3}$} \label{Fig2}
 \end{center}
\end{figure}

\textbf{Ising model}: The simplest example is the Ising model
$(A_{2},A_{3})$ since the model has diagonal modular invariant
partition function the fusion matrices of the boundary operators is
the same as the bulk case. The fusion graphs are as fig ~2 so the
boundary states are as follows
\begin{eqnarray}\label{cardyising}
|\mathbf{1}\rangle&=&\frac{1}{\sqrt{2}}|\mathbf{1}\rangle\rangle+\frac{1}{\sqrt{2}}|\epsilon\rangle\rangle+\frac{1}{\sqrt[4]{2}}|\sigma\rangle\rangle;\nonumber\\
|\epsilon \rangle&=&\frac{1}{\sqrt{2}}|\mathbf{1}\rangle\rangle+\frac{1}{\sqrt{2}}|\epsilon\rangle\rangle-\frac{1}{\sqrt[4]{2}}|\sigma\rangle\rangle;\nonumber\\
|\sigma \rangle&=&|\mathbf{1}\rangle\rangle-|\epsilon\rangle\rangle.
\end{eqnarray}
These equations reflect the $Z_{2}$ symmetry corresponding to
changing the sign of spin, this is also evident in the loop
representation; $n_{\epsilon}=n_{\mathbf{1}}=1$. Both operators give
$\kappa=3$, these loops are the domain walls between different
spins. It is worth mentioning that this symmetry comes from the
natural $Z_{2}$ symmetry of Dynkin diagram. The operator $\sigma$
with $n_{\sigma}=\sqrt{2}$ corresponds to free boundary condition.
The loops in the dense phase have $\kappa=\frac{16}{3}$ and describe
the domain walls of Fortuin-Kasteleyn (FK) clusters. In the above
calculation we considered only the largest eigenvalue of the fusion
graphs, however, it is also possible to consider other eigenvalues
as the weight of the loops, the cost is accepting complex local
Boltzmann weights for the corresponding height model. Since loop
models are generically non-local theories  accepting complex
Boltzmann weights is equal to accepting non-unitary theories. By
this introduction one can accept the possibility of loop models with
$n=\pm\sqrt{2},0$ for the loop model corresponding to the $A_{3}$
diagram of spin operator.
\newline

\textbf{Tri-critical Ising model}: The next simple example is the
tri-critical Ising model, $(A_{3},A_{4})$ which we have diagonal
modular invariant partition function. The boundary CFT of this model
was discussed in \cite{chim}. There are ~6 boundary operators
$\mathbf{1},\epsilon,\epsilon',\epsilon'',\sigma$ and $\sigma'$ with
the fusion graphs as fig ~3 and the following Cardy states
\begin{eqnarray}\label{cardytricritical}
|\mathbf{1}\rangle &=& C[|\mathbf{1}\rangle\rangle +
\eta|\epsilon\rangle\rangle + \eta|\epsilon'\rangle\rangle +
|\epsilon''\rangle\rangle + \root4\of{2}|\sigma'\rangle\rangle +
\root4\of{2}|\sigma\rangle\rangle];\nonumber\\
|\epsilon \rangle &=&C[\eta^2|\mathbf{1}\rangle\rangle -
\eta^{-1}|\epsilon\rangle\rangle - \eta^{-1}|\epsilon'\rangle\rangle
+ \eta^2|\epsilon''\rangle\rangle -
\root4\of{2}\eta^2|\sigma'\rangle\rangle +
\root4\of{2}\eta^{-1}|\sigma\rangle\rangle];\nonumber\\
|\epsilon'\rangle &=& C[\eta^2|\mathbf{1}\rangle\rangle -
\eta^{-1}|\epsilon\rangle\rangle - \eta^{-1}|\epsilon'\rangle\rangle
+ \eta^2|\epsilon''\rangle\rangle +
\root4\of{2}\eta^2|\sigma'\rangle\rangle -
\root4\of{2}\eta^{-1}|\sigma\rangle\rangle];\nonumber\\
|\epsilon''\rangle &=& C[|\mathbf{1}\rangle\rangle +
\eta|\epsilon\rangle\rangle + \eta|\epsilon'\rangle\rangle +
|\epsilon''\rangle\rangle - \root4\of{2}|\sigma'\rangle\rangle -
\root4\of{2}|\sigma\rangle\rangle];\nonumber\\
|\sigma'\rangle &=& \sqrt2C[|\mathbf{1}\rangle\rangle -
\eta|\epsilon\rangle\rangle + \eta|\epsilon'\rangle\rangle -
|\epsilon''\rangle\rangle] ;\nonumber\\
|\sigma\rangle &=& \sqrt2C[\eta^2|\mathbf{1}\rangle\rangle +
\eta^{-1}|\epsilon\rangle\rangle - \eta^{-1}|\epsilon'\rangle\rangle
- \eta^2|\epsilon''\rangle\rangle],
\end{eqnarray}
where $C = \sqrt{{sin{\pi\over5}}\over{\sqrt5}}$ and $\eta =
\sqrt{2\cos{\frac{\pi}{5}}}$. The boundary states corresponding to
boundary operators $\mathbf{1}$ and $\epsilon''$ can be transformed
to each other by just changing the sign of spin operators, i.e.
$Z_{2}$ symmetry. They have also the same loop weight $n=1$ comes
from the largest eigenvalue of the fusion matrix\footnote{To get the
loop weights we consider one simply connected part of the fusion
graph as an adjacency graph, the other parts of the graph have
always equal largest eigenvalues. One can see that these different
parts are folding or orbifold dual of each other, see
\cite{ginsparg} }. The boundary states $\epsilon$ and $\epsilon'$
are connected also by just changing the the sign of spin states. The
weight of the loops is $n=2\cos(\frac{\pi}{5})$ with $\kappa=5$ in
the dense phase. This loop model corresponds to the boundary of
geometric clusters at the geometric critical point of tri-critical
Ising model or Blume-Capel model ~\cite{blote nienhuis}. The
operator $\sigma'$ describes a loop model with $n=\sqrt{2}$
corresponding to $\kappa=\frac{16}{5}$ in the dense phase which is
related to the boundary of spin clusters and also vacancy clusters
in Blume-Capel model ~\cite{blote nienhuis,DGB}. The interesting
point for tri-critical models is the equality of critical exponents
for spin clusters and FK clusters ~\cite{blote nienhuis,DGB}.
\begin{figure}
\begin{center}
\includegraphics[angle=0,scale=0.6]{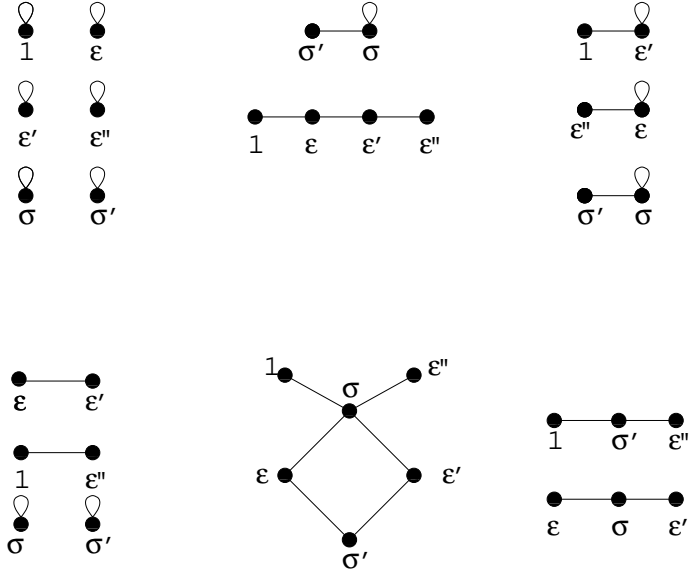}\\
\caption{Graphs of fusion matrices of the boundary primary operators
in the tri-critical Ising  model in the upper row from the left to
the right the fusion graphs of $1$, $\epsilon$ and $\epsilon'$, in
the lower row from the left to the right the fusion graphs of
$\epsilon''$, $\sigma$ and $\sigma'$. The fusion graph of $\epsilon$
is $A_{4}$ plus $T_{2}$, they are connected to each other by folding
duality. The fusion graph of $\sigma$ is $T_{2}\otimes A_{3}$.}
\label{Fig3}
 \end{center}
\end{figure}
The operator $\sigma$ is related to the degenerate boundary
condition and the corresponding loop model with
$n=2\sqrt{2}\cos(\frac{\pi}{5})$ is non-critical, however, it is
easy to see that the fusion matrix of this operator is decomposable
to simple matrices $N_{\sigma}=N_{T_{2}}\otimes N_{A_{3}}$ so one
can define for this graph two-flavor loop model with weights
$n_{1}=\sqrt{2}$ and $n_{2}=2\cos(\frac{\pi}{5})$. One can conclude
from the above discussion that those operators with the same loop
representations are connected to each other by folding and orbifold
duality and it is also possible to see these symmetries in the level
of boundary states.

Similar to the previous subsection one can also consider other
possible loop weights come from the other eigenvalues of the fusion
matrix. The eigenvalues of the fusion matrix of the operator
$\epsilon$ are
$n=\pm2\cos(\frac{\pi}{5}),\pm\frac{1}{2\cos(\frac{\pi}{5})}$ and
the eigenvalues of the fusion matrix of the operator $\sigma'$ are
$\pm\sqrt{2},0$. The eigenvalues of the other operators are a subset
of the above eigenvalues. Interestingly apart from the negative
eigenvalues the above weights can be fitted with the boundary loop
weights in \cite{Jacobsen and Saleur1,Jacobsen and Saleur2}.
\newline
\begin{figure}
\begin{center}
\includegraphics[angle=0,scale=0.6]{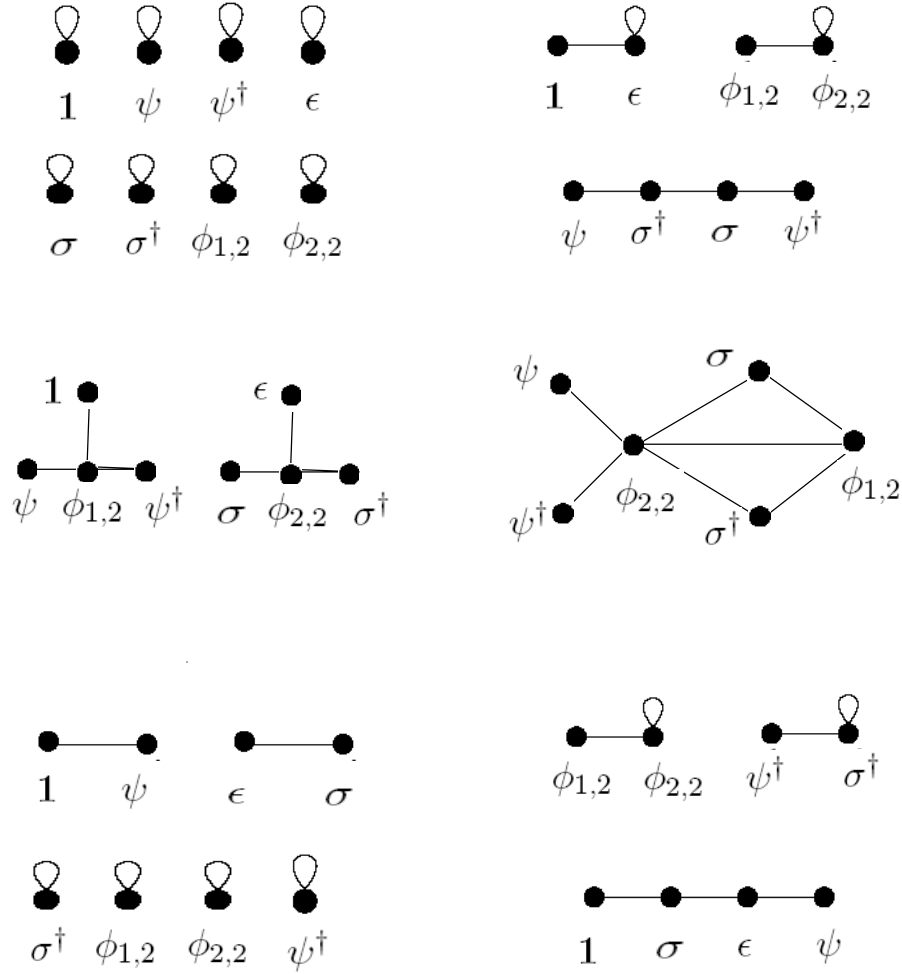}\\
\caption{Graphs of fusion matrices of primary operators in three
states Potts model. In the upper row from the left to the right the
fusion graphs of $1$ and $\epsilon$, in the middle row from the left
to the right the fusion graphs of $\phi_{1,2}$ and $\phi_{2,2}$ and
in the lowest row the fusion graphs of $\psi$ and $\sigma$. The
fusion graphs of $\psi^{\dag}$ and $\sigma^{\dag}$ can be derived
from the fusion graphs of $\psi$ and $\sigma$ by the following
exchanges $\psi\leftrightarrow \psi^{\dag}$ and
$\sigma\leftrightarrow \sigma^{\dag}$. The fusion graph of
$\epsilon$ is $A_{4}$ plus two $T_{2}$ graphs, they are connected to
each other by folding duality. The fusion graph of $\phi_{1,2}$ is
two $D_{4}$ and the fusion graph of $\phi_{2,2}$ is $T_{2}\otimes
D_{4}$.} \label{Fig3}
 \end{center}
\end{figure}

\textbf{Three states Potts model}: The next example is the first
non-diagonal case, ~3-state Potts model $(A_{4},D_{4})$ with ~8
boundary operators
$\mathbf{1},\psi,\psi^{\dag},\epsilon,\sigma,\sigma^{\dag},\phi_{1,2}$
and $\hat{\phi}_{2,2}$, see ~\cite{Cardy89, BPZ,saleur}. The fusion
graphs are given in fig ~4. Following Cardy's argument one can show
that the operators $\mathbf{1},\psi,\psi^{\dag}$ correspond to fix
boundary conditions and the corresponding boundary states can be
transformed to each other by $Z_{3}$ symmetry, i.e. the symmetry of
Dynkin diagram $D_{4}$. They also have  the same quantum dimensions
$n_{\mathbf{1}}=n_{\psi}=n_{\psi^{\dag}}=1$. The operators
$\epsilon,\sigma,\sigma^{\dag}$ describe the fluctuating boundary
conditions ~\cite{Cardy3} and all have the same kinds of fusion
graphs with
$n_{\epsilon}=n_{\sigma}=n_{\sigma^{\dag}}=2\cos(\frac{\pi}{5})$. In
the dilute phase one can consider $\kappa=\frac{10}{3}$ as the
property of the curve. In the lattice ~3-state Potts model these
loops are the same as the domain walls of spin clusters. The fusion
graph of the operator $\phi_{1,2}$ is two $D_{4}$ graphs. This
operator describes fix boundary condition and has loop model with
$n=\sqrt{3}$ which is equal to the loop model of domain walls in FK
clusters of ~3-state Potts model. The operator $\phi_{2,2}$
describes degenerate boundary condition and the corresponding loop
model with $n_{\phi_{2,2}}=\sqrt{\frac{9+3\sqrt{5}}{2}}$ is
non-critical, however, decomposition is possible. In this case one
can write $N_{\phi_{2,2}}=N_{T_{2}}\otimes N_{D_{4}}$ and so the
corresponding two-flavor loop model has weights $n_{1}=\sqrt{3}$ and
$n_{2}=2\cos(\frac{\pi}{5})$.

The fusion matrix of $\varepsilon$ as was discussed in the case of
tri-critical Ising model has the eigenvalues
$n=\pm2\cos(\frac{\pi}{5}),\pm\frac{1}{2\cos(\frac{\pi}{5})}$ and
the eigenvalues of the $N_{D_{4}}$ are $\pm\sqrt{3},0$. These loop
weights can be fitted with the boundary loop weights in
\cite{Jacobsen and Saleur1,Jacobsen and Saleur2}.
\newline
\newline
\textbf{Tri-critical three states Potts model}: The next interesting
example is tri-critical ~3-state Potts model $(D_{4}, A_{6})$  it
has non-diagonal modular invariant partition function and also it is
not part of Pasquirer's A-D-E  models. The boundary states of this
model have not been investigated systematically so far. The boundary
operators of this model are: $\phi_{i}$ with $i=(r,a)$, $r=1,2,3$
and $a=1,...,4$. The fusion graphs for the boundary operators in
this case are given in the Appendix. The boundary states
corresponding to boundary operators
$\phi_{1,1},\phi_{1,3},\phi_{1,4}$ can be transformed to each other
by $Z_{3}$ symmetry of spin operators and should correspond to fix
boundary conditions with $n=1$. The operators
$\phi_{2,1},\phi_{2,3},\phi_{2,4}$ have also the same property with
the same fusion graphs with $n=2\cos(\frac{\pi}{7})$. In the lattice
tri-critical ~3-state Potts model they are domain walls of geometric
clusters of geometric critical point ~\cite{blote nienhuis} with
$\kappa=4\frac{7}{6}$. The operators
$\phi_{3,1},\phi_{3,3},\phi_{3,4}$ can be transformed to each other
again by $Z_{3}$ symmetry but they have loop weights bigger than
two; $n=2.246$. The operators $\phi_{2,2}$ and $\phi_{3,2}$ have
also loop weights bigger than two and related to degenerate boundary
conditions. Finally the graph of $\phi_{1,2}$ is equal to three
$D_{4}$ graphs with $n=\sqrt{3}$.  In the dilute phase this weight
describes the domain walls of spin clusters in the lattice
tri-critical ~3-state Potts model with $\kappa=4\frac{6}{7}$.

The fusion graph of $\phi_{2,1}$ is the sum of two graphs $A_{6}$
and $T_{3}$. The fusion matrix has the eigenvalues
$n=2\cos(\frac{\pi j}{7})$ with $j=2,3,4,5,6$. The eigenvalues of
the fusion matrix of $\phi_{1,2}$ are $n=\pm\sqrt{3},0$.
Interestingly again apart from the negative eigenvalues the above
weights can be fitted with the boundary loop weights in
\cite{Jacobsen and Saleur1,Jacobsen and Saleur2}. The fusion graph
of $\phi_{2,2}$ is decomposable as $T_{3}\otimes D_{4}$ and so it is
possible to define two crossing loop models in this case. The fusion
graphs of $\phi_{3,1}$ is not decomposable to simple graphs so it is
not possible to extract critical loops also for $\phi_{3,3}$ and
$\phi_{3,4}$ which are in the same sector. Although the loops,
extracted by our method, corresponding to the above operators are
not critical but by considering the fusion graph of the ground state
of the above adjacency graph it is possible to extract critical
loops. we will not discuss this method here, for more detail one can
see \cite{Fendley}. The fusion graph of $\phi_{3,2}$ is decomposable
but not to the simple graphs, i.e.
$N_{\phi_{3,2}}=N_{T_{3}^{2}}\otimes N_{D_{4}}$. Another possibility
to extract critical loops for $\phi_{3,1}$ is by considering other
eigenvalues of the fusion matrix of this operator. The eigenvalues
of $N_{\phi_{3,1}}$ are
$\pm\frac{\sin(\frac{3\pi}{7})}{\sin(\frac{\pi}{7})}$,
$\pm\frac{\sin(\frac{2\pi}{7})}{\sin(\frac{\pi}{7})}$ and
$\pm\frac{\sin(\frac{\pi}{7})}{\sin(\frac{2\pi}{7})}$, the last two
cases have critical loops.
\newline
\newline
\textbf{Minimal models}: Finding loop models by the above method is
completely general and applicable for more general cases. Take a
pair $(A,G)$ from the equation (\ref{ADE}) then it is possible to
correspond at least two different kinds of loop models for these
minimal models with the following weights
\begin{eqnarray}\label{loop minimal}
n=2\cos(\frac{\pi}{g}),\hspace{2cm}n=2\cos(\frac{\pi}{h}).
\end{eqnarray}
They are the largest eigenvalues of the fusion matrices of
$\phi_{1,2}$ and $\phi_{2,1}$. One can also consider the following
SLE drifts for these loop models
\begin{eqnarray}\label{SLE}
\kappa=4\frac{g}{g+1},\hspace{2cm}\kappa=4\frac{h}{h-1},\hspace{1cm}g-h&=&1,\nonumber\\
\kappa=4\frac{g}{g-1},\hspace{2cm}\kappa=4\frac{h}{h+1},\hspace{1cm}g-h&=&-1.
\end{eqnarray}
The other eigenvalues of $G$ can be written as
\begin{eqnarray}\label{exponents}
n=2\cos(\frac{\pi m}{g}),
\end{eqnarray}
where $m$ is one of the Coexter exponents of the graph $G$. they are
listed in the table ~1.

It is possible to consider loop models for the above eigenvalues as
before, however, they are not still all the possible loop models
because as we already showed in some cases one can define two flavor
loop models for decomposable fusion graphs. It is also possible as
the case of the fusion graph of $\phi_{3,1}$ in tri-critical
~3-state Potts model to have matrices with relevant non-largest
eigenvalues. We believe that they are relevant because the same loop
weights appear in the classification of Jacobsen and Saleur
\cite{Jacobsen and Saleur1}.

 Although so far we have given more familiar examples
as the possible candidates for our loop models but it is also
possible to extract systematic examples for the above proposals by
using Pasquier's ADE models and Dilute ADE models \cite{Roc92,WN}.
Pasquier's ADE models give a lattice realization for the $(A,G)$
series with $g-h=1$ and the description briefly is as follows:
define an RSOS model by using the graph $G$ this height model at the
critical point can be described by a the minimal CFT then map this
height model to loop model \cite{Cardy ADE} at the critical point
with $n=2 \cos(\frac{\pi}{g})$ which is the same as the loop model
that we proposed in (\ref{loop minimal}). Of course the method
proposed in this article and \cite{Rajabpour} is highly influenced
with Pasquier's ADE models but it has something more to say by
connecting the loop properties to the fusion properties of the
primary operators.
\begin{figure}
\begin{center}
\includegraphics[angle=0,scale=0.5]{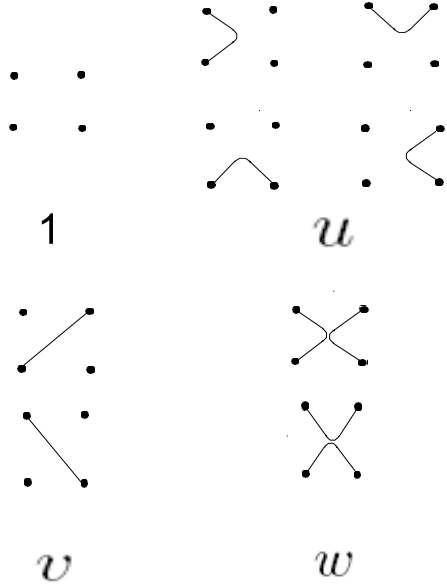}\\
\caption{The Boltzmann weights of the different vertices in the
$O(n)$ model on the square lattice.} \label{Fig1}
 \end{center}
\end{figure}
To get the dilute loop models and the loop models corresponding to
tri-critical models we need to use Dilute ADE models. These models
have rich phase diagrams with four branches: branch ~1 and ~2 have
central charges $c=1-\frac{6}{g(g\pm1)}$ and branch ~3 and ~4 have
$c=\frac{3}{2}-\frac{6}{g(g\pm1)}$. One can also map this height
models to $O(n)$ loop models with the non-intersecting bonds on the
square lattice with the partition function
\begin{eqnarray}\label{partition function of generalized loop model}
Z=\sum u^{N_{u}}v^{N_{v}}w^{N_{w}}n^{N},
\end{eqnarray}
where the weights for different plaquettes are given in Fig ~5 and
the $N_{u},N_{v}$ and $N_{w}$ are the numbers of different
plaquettes \cite{BN}. This generalized $O(n)$ loop model apart from
the critical properties at $u=w=\frac{1}{2}$ and $v=0$ has four
other branches coincide with the four branches of dilute ADE models
\cite{warnaar}. The weights are given by
\begin{eqnarray}\label{weights}
n&=&-2\cos(2\theta),\nonumber\\
w&=&\frac{1}{2-[1-2\sin(\theta/2)][1+2\sin(\theta/2)]^{2}},\nonumber\\
u&=&\pm 4w \sin(\theta/2)\cos( \pi /4-\theta /4),\\
v&=&\pm w[1+2\sin(\theta/2)],\nonumber
\end{eqnarray}
where $\frac{\pi}{2}\leq\theta\leq\pi$,
$0\leq\theta\leq\frac{\pi}{2}$, $-\frac{\pi}{2}\leq\theta\leq0$ and
$-\pi\leq\theta\leq-\frac{\pi}{2}$ are the intervals corresponding
to branches ~1, ~2, ~3 and ~4 respectively. They coincide with the
different branches in the dilute ADE models.

 It is interesting to
investigate the connection of the above loop model to the SLE. There
are different methods to do that here we use the magnetic operator
to find the SLE drift. It was shown in \cite{BN} by the numerical
calculation that the magnetic exponent of the branch ~1 and ~2 is
identified with $2h_{\frac{m+1}{2},\frac{m+1}{2}}$ where

\begin{eqnarray}\label{Kac formula}
h_{r,s}=\frac{((m+1)r-ms)^{2}-1}{4m(m+1)},
\end{eqnarray}
and $m$ is related to the central charge of the theory by
$c=1-\frac{1}{m(m+1)}$. Its connection to the loop variables comes
from the relation
$\frac{2\theta}{\pi}+\frac{\pi}{2\theta}-2=\frac{1}{m(m+1)}$ derived
from the coulomb gas method \cite{BN}. The connection of the
magnetic exponent to the SLE drift is as follows \cite{Sheffield}
\begin{eqnarray}\label{magnetic exponentand kappa}
2h_{\frac{m+1}{2},\frac{m+1}{2}}=\frac{(8-\kappa)(3\kappa-8)}{32\kappa}.
\end{eqnarray}
Using the above equation the SLE drift at the branches ~1 and ~2 of
the loop model (\ref{partition function of generalized loop model})
can be derived as follows
\begin{eqnarray}\label{SLE drift}
\kappa&=&\frac{8\theta}{\pi}.
\end{eqnarray}
This result is consistent also with our expectation from the second
level null vector of minimal models \cite{bernard}, it is also
consistent with the recent direct investigation by using holomorphic
variables \cite{cardy4} .

Back to the height model representation one can summarize following
results: the branch ~2 of the ADE models corresponds to the dilute
loops with $n=2 \cos(\frac{\pi}{h})$ and the branch ~1 is the dense
phase of tri-critical models with $n=2 \cos(\frac{\pi}{h})$. The
results for some of the simple cases are as follows:
\begin{eqnarray}\label{ll}
\mbox{branch 2:}\quad A_2&=&\mbox{critical percolation,}\hspace{1.5cm}c=0 \hspace{1.5cm} n=1;\nonumber\\
\mbox{branch 1:}\quad A_2&=&\mbox{critical Ising}\hspace{2.6cm}c=1/2\hspace{1.1cm}  n=1;\nonumber\\
\mbox{branch 2:}\quad A_3&=&\mbox{critical Ising}\hspace{2.6cm}c=1/2\hspace{1.1cm} n=\sqrt{2};\nonumber\\
\mbox{branch 1:}\quad A_3&=&\mbox{tri-critical Ising}\hspace{2.1cm}c=7/10\hspace{.9cm} n=\sqrt{2};\nonumber\\
\mbox{branch 2:}\quad A_4&=&\mbox{tri-critical Ising}\hspace{2.1cm}c=7/10\hspace{.92cm} n=2\cos(\frac{\pi}{5});\nonumber\\
\mbox{branch 2:}\quad D_4&=&\mbox{critical 3-state Potts}\hspace{1.2cm}c=4/5\hspace{1.1cm} n=\sqrt{3};\nonumber\\
\mbox{branch 1:}\quad D_4&=&\mbox{tri-critical 3-state
Potts}\hspace{0.8cm}c=6/7\hspace{1.1cm} n=\sqrt{3}.\nonumber
\end{eqnarray}

Using the above method it is easy to find the lattice realization
for most of the proposed loop models, the results are interestingly
consistent. Following the same method it is possible to extract the
loop models corresponding to minimal CFTs, however, the loop model
for the non-diagonal cases with $g-h=-1$ is not extractable with
this method because we are not able to find the dense phase of loop
models for these cases. It seems that the dense lattice height model
has not been proposed for this case.

To conclude this subsection we proposed some loop representations
for the minimal CFTs by using fusion of boundary operators. Then
since ADE models give a lattice statistical model representation for
minimal CFTs we used these models to extract physical loop models
corresponding to ADE models. The fractal properties of these lattice
loop models are the same as the loop models that we proposed by
using the fusion of primary operators.
\newline
\newline
 \textbf{$SU_{k}(2)$ Models}: It is possible to follow the same
calculation for every unitary minimal model. For example for WZW
$SU_{k}(2)$ models the classification of modular invariant partition
functions is based on A-D-E-T graphs with $g=k+2$. The same method
as the minimal models is applicable here and one can find boundary
operators $\hat{\phi}_{j}$ with $1\leq j\leq k+1$. The loop models
have weights $d_{j}=\frac{\sin(\frac{\pi
j}{g})}{\sin(\frac{\pi}{g})}$. Only $j=\frac{1}{2}$ has critical
loop representation with the following loop weight
\begin{eqnarray}\label{loop weight}
n=2\cos(\frac{\pi}{k+2}),
\end{eqnarray}
with $\kappa=4\frac{k+2}{k+3}$ and $\kappa=4\frac{k+2}{k+1}$ for the
dilute and dense phase respectively. The other loop models are not
critical except for $k=4$ with $n=2$. The fusion graphs of the
operators with $j\neq \frac{1}{2}$ is not decomposable to the simple
graphs, however, the non-largest eigenvalues can be still relevant.
For example take $k=5$ with $j=3/2$, the fusion graph is similar to
the one part of the $\phi_{31}$ fusion graph of the tri-critical
~3-state Potts model, the right one in fig ~6. The eigenvalues are
$\pm\frac{\sin(\frac{3\pi}{7})}{\sin(\frac{\pi}{7})}$,
$\pm\frac{\sin(\frac{2\pi}{7})}{\sin(\frac{\pi}{7})}$ and
$\pm\frac{\sin(\frac{\pi}{7})}{\sin(\frac{2\pi}{7})}$, the last two
cases have critical loop representation. The similarities between
fusion graphs of $SU_{k}(2)$ models with minimal models is not just
an accident they are based on the coset construction of the minimal
models.

\section{Discussion}\

We proposed a method to classify some possible loop models
consistent with the conformal boundary conditions for generic
rational CFT: take the simply laced classification of the
corresponding minimal CFT then find the boundary operators and also
the fusion matrices, make the $O(n)$ loop model of the primary
operator by the method that we discussed in section ~3 and
~\cite{Rajabpour}. We think that there should be some connections
between these loop models and the SLE interpretation of CFT
investigated in \cite{bernard} which is based on the connection of
SLE with the null vectors in the CFT. This connection is not
complete even for minimal CFTs because we do not know how to explain
the boundary operators with the same loop model but with the
different null vectors, for example in the three states Potts model
$\epsilon,\sigma$ and $\sigma^{\dag}$ are in the same sector from
the boundary CFT point of view but just $\epsilon$ and $\sigma$ have
the required second level null vectors. However, from null vector
point of view this correspondence is not clear but it is possible to
show that in the partition function level this similarity is more
known. Another way to look at the results of this paper is by
conjecturing the largest eigenvalue of the fusion graph as the
possible loop weight for the loop model in the universality class of
the corresponding CFT without defining any height model on the
fusion graph.

One possible generalization of the above construction is by
considering graphs with largest eigenvalue bigger than ~2 as an
adjacency graph of fused RSOS model and then extracting the loop
model by the method investigated in \cite{Fendley}. The other
interesting direction is to investigate the modular invariant
partition functions of loop models and their possible connections to
the classified modular invariant partition functions of minimal
models, this is related to investigate more directly the connection
of our method to the classification of \cite{Jacobsen and
Saleur1,Jacobsen and Saleur2}.
\newline
\newline

 \textit{ Acknowledgment}:

I thank Roberto Tateo for useful discussions and Paul Fendley for
useful comments. I thank also John Cardy for his useful criticism.

\section{Appendix}\
\begin{figure}
\begin{center}
\includegraphics[angle=0,scale=0.6]{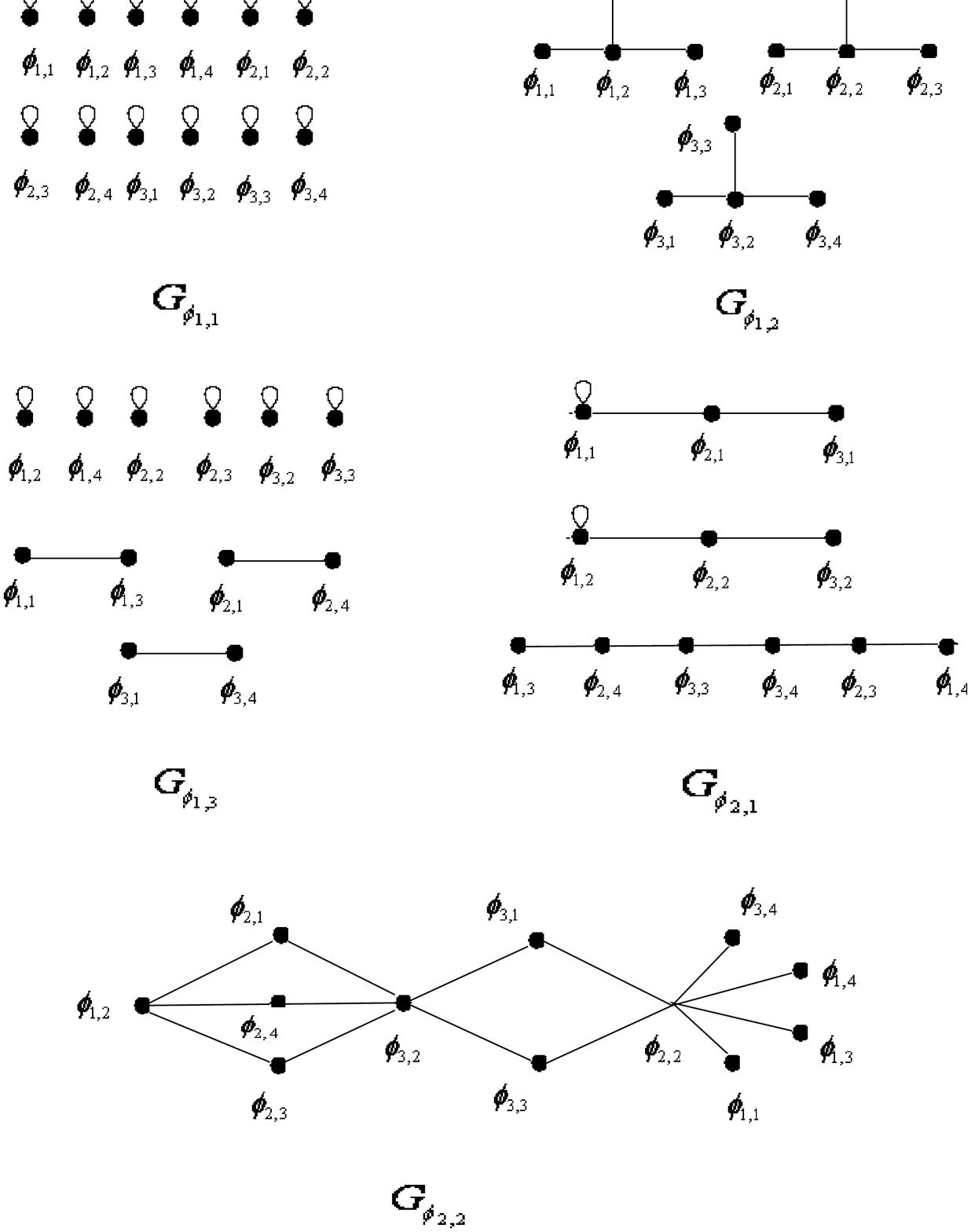}\\
 \end{center}
\end{figure}
\begin{figure}
\begin{center}
\includegraphics[angle=0,scale=0.6]{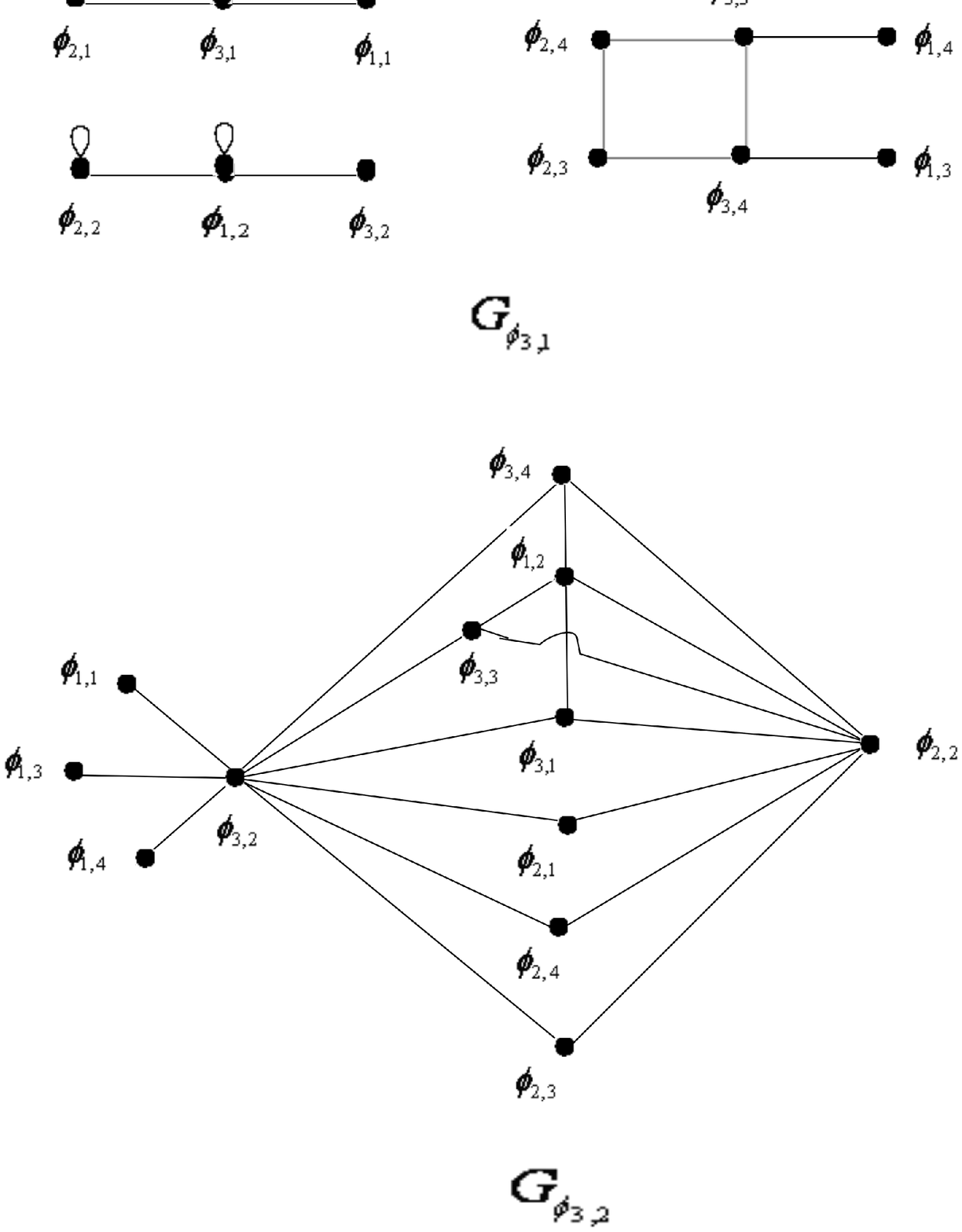}\\
 \end{center}
\end{figure}

In this appendix we list the fusion graphs of the boundary operators
in tri-critical ~3-state Potts model. The fusion graphs are given in
fig ~6. The fusion graph of $\phi_{1,4}$ can be derived from the
fusion graph of the operator $\phi_{1,3}$ by the following
transformations:
\begin{eqnarray}\label{transformations1}
\phi_{1,3}\leftrightarrow
\phi_{1,4}\hspace{1cm}\phi_{2,3}\leftrightarrow
\phi_{2,4}\hspace{1cm}\phi_{3,4}\leftrightarrow \phi_{3,3}.
\end{eqnarray}
The fusion graph of $\phi_{2,3}$ can be derived from the fusion
graph of the operator $\phi_{2,1}$ by the following transformations:
\begin{eqnarray}\label{transformations2}
\phi_{1,3}\leftrightarrow
\phi_{1,1}\hspace{1cm}\phi_{2,3}\leftrightarrow
\phi_{2,1}\hspace{1cm}\phi_{3,3}\leftrightarrow \phi_{3,1}.
\end{eqnarray}
Finally the fusion graph of $\phi_{2,4}$ can be derived from the
fusion graph of the operator $\phi_{2,1}$ by the following
transformations:
\begin{eqnarray}\label{transformations3}
\phi_{1,4}\leftrightarrow
\phi_{1,1}\hspace{1cm}\phi_{2,4}\leftrightarrow
\phi_{2,1}\hspace{1cm}\phi_{3,4}\leftrightarrow \phi_{3,1}.
\end{eqnarray}
To get the fusion graphs of $\phi_{3,3}$ and $\phi_{3,4}$ from the
fusion graph of $\phi_{3,1}$ one just need to use the
transformations (\ref{transformations2}) and
(\ref{transformations3}) respectively.

We shall call the part of the fusion graph of $\phi_{3,1}$ with two
neighbor blobs $T_{3}^{2}$, the lower index is the number of nodes
and the upper index is the number of blobs attached to the
neighboring nodes of the graphs starting from one of the extremes.
These kinds of fusion graphs appear also in the fusion graph of
$\phi_{j=1}$ of $SU_{2}(k)$ models.

\end{document}